\documentclass[prb,twocolumn,a4paper,floatfix,superscriptaddress,showpacs,showkeys]{revtex4-1}

\usepackage{dcolumn,amsmath,xspace}
\usepackage{graphicx}
\usepackage{subfigure}
\usepackage{amssymb}

\usepackage{epstopdf}

\newcommand{\degC}{\ensuremath{^{\circ}\text{C }}}

\begin{document}
\title{Wide-gap Semiconducting Graphene from Nitrogen-seeded SiC \\
}

\author{F. Wang}
\affiliation{School of Physics, The Georgia Institute of Technology, Atlanta, Georgia 30332-0430, USA}
\author{G. Liu}
\affiliation{Institute for Advanced Materials Devices and Nanotechnology, Rutgers University, Piscataway, New Jersey 08854}
\author{S. Rothwell}
\affiliation{Department of Electrical and Computer Engineering, University of Minnesota, Minneapolis, Minnesota 55455}
\author{M. Nevius}
\affiliation{School of Physics, The Georgia Institute of Technology, Atlanta, Georgia 30332-0430, USA}
\author{A. Tejeda}
\affiliation{Institut Jean Lamour, CNRS - Univ. de Nancy - UPV-Metz, 54506 Vandoeuvre les Nancy, France}
\affiliation{Synchrotron SOLEIL, L'Orme des Merisiers, Saint-Aubin, 91192 Gif sur Yvette, France}
\author{A. Taleb-Ibrahimi}
\affiliation{UR1 CNRS/Synchrotron SOLEIL, Saint-Aubin, 91192 Gif sur Yvette, France}
\author{L.C. Feldman}
\affiliation{Institute for Advanced Materials Devices and Nanotechnology, Rutgers University, Piscataway, New Jersey 08854}
\author{P.I. Cohen}
\affiliation{Department of Electrical and Computer Engineering, University of Minnesota, Minneapolis, Minnesota 55455}
\author{E.H. Conrad}
\affiliation{School of Physics, The Georgia Institute of Technology, Atlanta, Georgia 30332-0430, USA}

\begin{abstract}
\textbf{All carbon electronics based on graphene has been an elusive goal.  For more than a decade, the inability to produce significant band-gaps in this material has prevented the development of graphene electronics.  We demonstrate a new approach to produce semiconducting graphene that uses a submonolayer concentration of nitrogen on SiC sufficient to pin epitaxial graphene to the SiC interface as it grows.  The resulting buckled graphene opens a band-gap greater than 0.7eV in the otherwise continuous metallic graphene sheet.
}
\end{abstract}
\vspace*{4ex}

\maketitle
\newpage


The goal of developing all-carbon electronics requires the ability to dope graphene and convert it between metallic and wide band-gap semiconducting forms.  While doping graphene by adsorbates or more elaborate chemical means has made rapid progress,\cite{Kubler_PRB_05,Gierz_NatM_08,Coletti_PRB_10,Emtsev_PRB_11,N_review,Fort_ACSNano_12,Usachov_NanoLett_11} opening a band-gap in graphene has been problematic. Two routes to wide-band-gap semiconducting graphene have been pioneered: electron confinement and chemical functionalization. Electron confinement in lithographically patterned narrow graphene ribbons has been plagued by lithographic limits and edge disorder,\cite{Melinda_PRL_10,jiao2009,jiao2010,Oostinga_PRB_10}  although recent results from sidewall grown graphene ribbon are showing new progress that could lead to band-gaps larger than 0.6eV.\cite{Hicks_NatureP_13,Baringhaus_Ballistic}  Functionalized graphene band-gaps can be produced by imposing a periodic potential in the graphene lattice through ordered adsorbates\cite{Haddon_NLet_10,Balog_NMat_10} or ordered impurities replacing carbon atoms.\cite{Deifallah_JPHYSCHEM_08}

In this work we demonstrate a novel approach to band-gap engineering in graphene using a nitrogen seeded SiC surface.  Rather than using chemical vapor deposition (CVD) or plasma techniques to dope graphene by post seeding the films with nitrogen,\cite{Usachov_NanoLett_11,N_review,Joucken_PRB_12,Fort_ACSNano_12} we show that a submonolayer concentration of nitrogen adsorbed on SiC, prior to graphene growth, causes a large band-gap to open in the subsequently grown, continuous graphene sheets.  Using X-ray photoemission spectroscopy (XPS), scanning tunneling microscopy (STM), and angle resolved photoemission spectroscopy (ARPES), we show that a
submonolayer concentration of bonded nitrogen at the graphene-SiC interface leads to a 0.7eV semiconducting form of graphene.

The band-gap is not due to chemical functionalization since the concentrations used in these studies are expected to have little effect on graphene's band structure.\cite{Lherbier_PRL_08,Deifallah_JPHYSCHEM_08}  Instead, STM topographs and $dI/dV$ images, showing that the graphene is buckled into folds with 1-2nm radii of curvature, suggests two possible origins for the gap: either a quasi-periodic strain\cite{Low_PRB_11} or electron localization in the 1-2 nm wide buckled ribbons.\cite{Wakabayashi_STAM_10}

\begin{figure}
\includegraphics[width=8cm,clip]{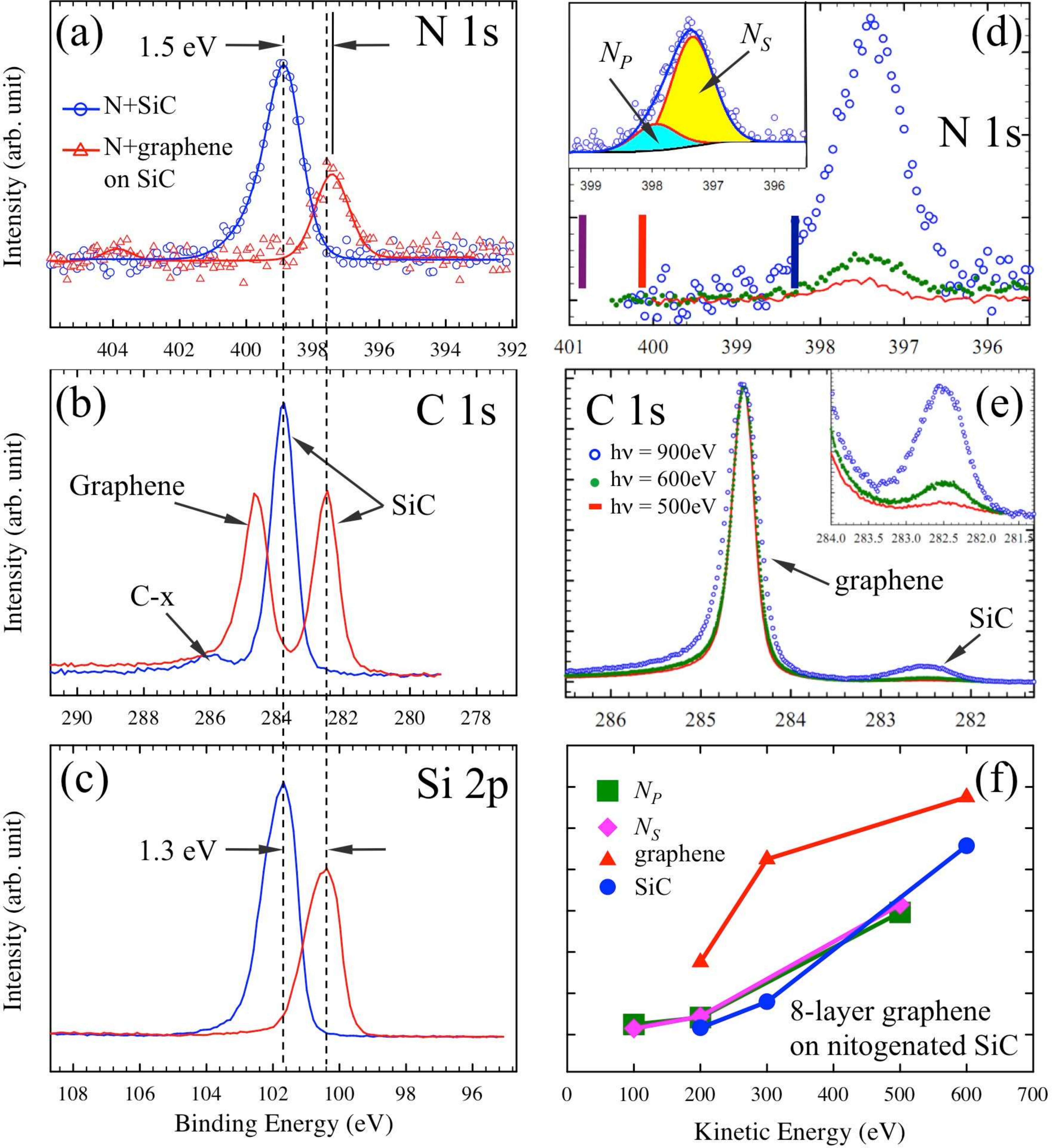}
\caption{(a)-(c) XPS from SiC$(000\bar{1})$ surface before and after growth of a 3-layer graphene film.  The pre- and post-growth N coverage is 0.3ML and 0.2ML, respectively  (a) N1s, (b) C1s, and (c) Si2p [$h\nu\!=\!1486$eV]. A 1.3eV shift after graphene forms is shown (dashed line).  The C-x peak at 286eV is from ``adventitious carbon" contamination in the initial surface.\cite{Avila_JAP_01} (d) and (e) XPS spectra for an 8-layer graphene film grown from the pre-nitrogen surface using photon energies of 500, 600, and 900eV.  The intensities have been normalized to the graphene C1s peak.  (d) N1s spectra at the same photon energies as (e). Insert shows the 2-component fit to the N1s peak.  The purple, red and blue bars in (d) mark the expected peak positions of pyrrolic, graphitic and pyridinic N-inclusions in graphene, respectively.\cite{N_review,Joucken_PRB_12}   (e) XPS of the C1s spectrum for the same photon energies in (d). Insert in (e) shows the SiC portion of the C1s XPS spectra. (f) The trend in photoelectron intensity versus photoelectron KE from an 8-layer graphene sample for the SiC C1s, graphene C1s, and the two N1s XPS peaks. Absolute intensities have been scaled by arbitrary factors to highlight the energy trend for each peak.} \label{F:STABLE}
\end{figure}

To produce this semiconducting form of graphene, we pre-deposit nitrogen onto a SiC$(000\bar{1})$ surface [see experimental details and supplemental material].  After nitrogenation, there is a strong N1s peak at 398.9eV [see Fig.~\ref{F:STABLE}(a)]
 with a FWHM=1.4eV.  The N1s binding energy (BE) and the broad width of the peak are consistent with a range of calculated sites that are likely due to two- and three-coordinated nitrogen in substitutional carbon sites bonded to interfacial silicon and carbon atoms [see supplemental material].\cite{Snis_PRB_99} In these studies the integrated N1s XPS intensity corresponds to starting nitrogen concentration of $\sim\!0.3$ML  ($\sim\!3.9\!\times\!10^{14}\text{cm}^{-2}$ of N on the 4H-SiC($000\bar{1}$) surface). The XPS data of Fig.~\ref{F:STABLE} shows that a large fraction of this N remains at the SiC-graphene interface after graphene growth at 1450\degC in the controlled silicon sublimation (CSS) furnace [see supplemental material].\cite{WaltPNAS} We suggest a model where a portion of the remaining nitrogen is not incorporated into the graphene but instead pins the graphene to the SiC to produce buckled semiconducting graphene.


 After growth, the graphene C1s peak develops at 284.5 eV [see Fig.~\ref{F:STABLE}(b)].  No oxides (C-O or Si-O) are measurable once the graphene has formed.  We note that both the Si 2p and SiC C1s peaks shift 1.3 eV to lower BE once the graphene forms [see Figs.~\ref{F:STABLE}(a) and (c)], consistent with the known band bending on graphitized C-face SiC.\cite{Seyller_MSF_07,Emtsev_PRB_08} Furthermore, the C 1s is essentially identical to clean C-face graphene (see for example Ref.~[\onlinecite{Emtsev_PRB_08}]).  After growth, the N1s peak is shifted by 1.5eV to lower BE (a net shift of -0.2 eV) and the total nitrogen concentration, as determined by XPS, reduces to $\sim\!2\!\times\!10^{14}/\text{cm}^{2}$, $\sim\!0.2$ML [see Experimental Methods]. The post growth N1s peak can be fit with two narrow pseudo-Voight peaks at 398.0 eV and 397.4eV (with FWHM of 0.8 and 0.9eV), labeled as $N_P$ and $N_S$, respectively [see the insert in Fig.~\ref{F:STABLE}(d)].  It is important to note that there are no other higher BE N1s peaks in the spectrum that would normally be associated with common nitrogen incorporation sites in the graphene lattice.\cite{N_review} In other words, little if any nitrogen is incorporated into the graphene lattice.

\begin{figure*}
\includegraphics[width=15.0cm,clip]{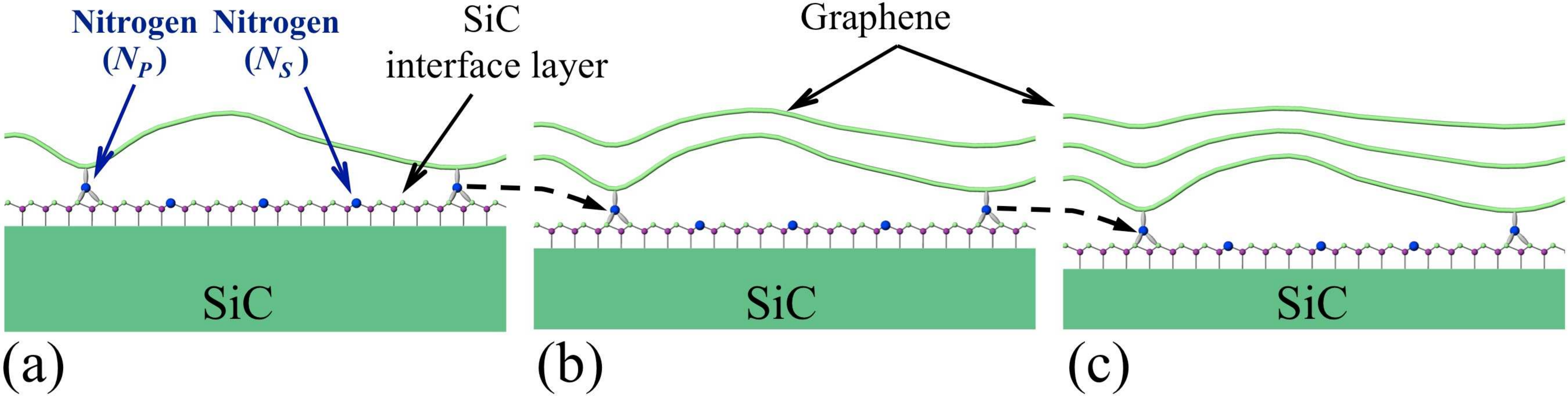}
\caption{Schematics showing how graphene layers grow from nitrogen-seeded SiC. (a) Most of the nitrogen, $N_S$, is bonded to Si atoms (magenta) in carbons sites. The remaining interfacial nitrogen, $N_P$, is bonded to both the SiC carbon atoms (green) and the graphene, pinning it to the SiC.
As the graphene grows and the SiC recedes, nitrogen remains at the SiC interface. (b) and (c) both show that the buckling amplitude in layers farther from the surface is reduced as strain is relieved.} \label{F:N2_Growth}
\end{figure*}

The dominant peak at 397.4 eV, labelled as Ns, is usually associated with N-SiC bonds.  Specifically, nitrogen in carbon sites bonded to Si atoms at the SiC interface as shown in Fig.~\ref{F:N2_Growth}(a)  [see supplemental material].\cite{Tochihara_ProgSS_11} The weak $\text{N}_P$ peak at 398.0 eV ($\sim\!25\%$ of the total nitrogen coverage) is in the energy range associated with pyridinic nitrogen sites in graphene,\cite{Shimoyama_PRB_00,Snis_PRB_99,Usachov_NanoLett_11,Fort_ACSNano_12} but can also be associated with other $\text{sp}^3$ and $\text{sp}^2$ C-N bonds in carbon nitride films.\cite{Ronning_PRB_98}  Photon energy-dependent XPS measurements in Figs.~\ref{F:STABLE}(d),(e), and (f) and STM images described below indicate that $\text{N}_P$ is not pyridinic nitrogen but is instead a second C-N compound at the graphene-SiC interface that pins the graphene to the SiC [see Fig.~\ref{F:N2_Growth}(a)].


Figures \ref{F:STABLE}(d) and (e) show XPS spectra of N1s and C1s for different photon energies.  A photon energy of $h\nu=500$eV produces photoelectrons with the shortest mean free path of the three photon energies used in these experiments and is therefore the most surface sensitive of the three.  In Fig.~\ref{F:STABLE}(e) the more surface sensitive spectra have the weakest SiC C1s, indicating that the SiC spectra is attenuated by the graphene layers above. The N1s peak is similarly weakest at this energy indicating that most of the nitrogen is near the SiC surface.  We also note that the ratio of the $N_S$ and $N_P$ peak intensities is independent of photon energy.  This is demonstrated in Fig.~\ref{F:STABLE}(f) where the intensity of the two peaks have essentially the same kinetic energy (KE) dependence.
These two findings confirm that the $N_P$ peak is associated with a nitrogen site at the SiC-graphene interface and is \emph{not} a pyridinic nitrogen inclusion in graphene.

The XPS results shown in Fig.~\ref{F:STABLE}(a) make clear that while a significant portion of the nitrogen desorbs during the high temperature (1450\degC) growth, the remaining nitrogen is maintained in a stable concentration of sites at the graphene-SiC interface. This implies that any high energy substitutional nitrogen sites that develop in the growing graphene film are annealed out and remain bonded at more favorable sites at the SiC interface.  Assuming a uniform distribution of $N_S$ and $N_P$  sites, the average distance between nitrogen atoms is $\sim\!0.9$nm for $N_S$ sites and $\sim\!1.5$nm for $N_P$ sites.  A length scale comparable to the  $N_P$ average spacing will be seen again in STM results from these same films and will be important in understanding the band structure of this material.

 \begin{figure*}
\includegraphics[width=15.0cm,clip]{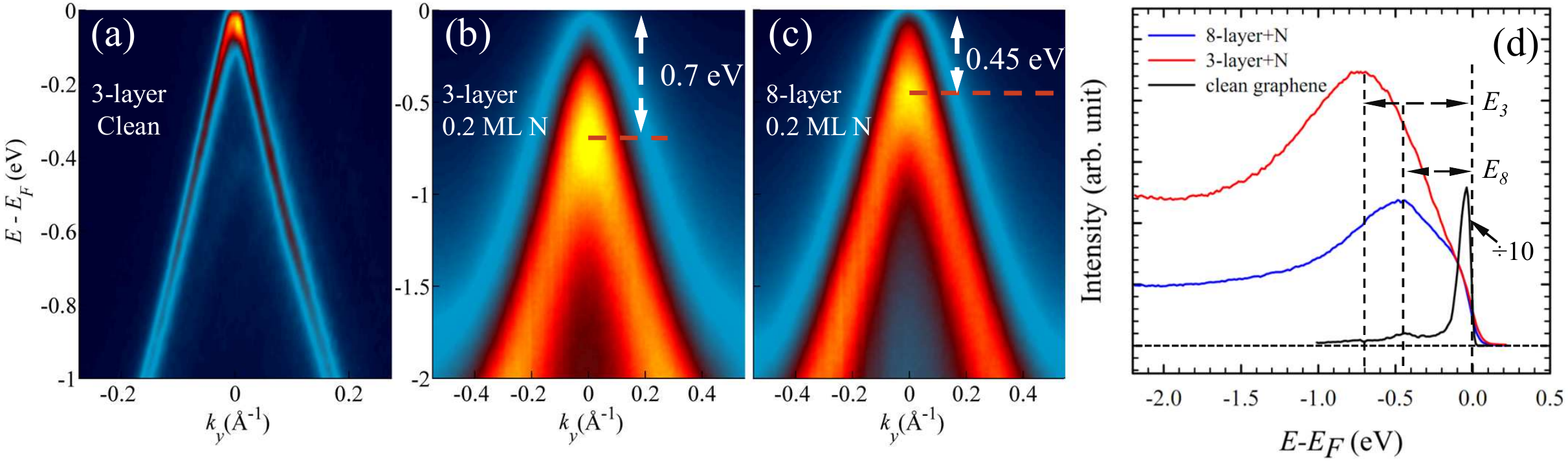}
\caption{The ARPES band structure taken with $\hbar \omega\!=\!36$ eV near the $K$-point of (a) clean 3-layer graphene (T=100K), (b) 3-layer graphene with a 0.2ML nitrogen SiC$(000\bar{1})$ surface (T=300K),  and (c) 8-layer graphene from a 0.2ML nitrogen SiC$(000\bar{1})$ surface (T=300K). $k_y$ is perpendicular to the $\Gamma K$ direction.  Note the different energy and momentum scales for (b) and (c) compared to (a). (d) Energy distribution curves through the Dirac point in (a), (b) and (c) show the 0.7eV gap for 3-layer films that reduces to 0.45eV for 8-layer films. Note that the apparent energy broadening observed near $E_F$ in the nitrogen samples compared to the clean surface, is entirely due to the temperature difference between the two measurements that broadens the 300K Fermi-Dirac function.} \label{F:ARPES}
\end{figure*}

 The interfacial N-preparation method used here causes a band-gap to open in graphene's $\pi$ bands.  We demonstrate this using high resolution ARPES from graphene films taken at the graphene $K$-point (rotated $30^\circ$ from the SiC $<\!10\bar{1}0\!>$ direction).  Figure \ref{F:ARPES}(a) shows the typical band dispersion perpendicular to the $\Gamma K$ direction at the graphene $K$-point from a clean 3-layer film.  As the ARPES shows, pristine 3-layer C-face graphene consists of linear $\pi$-bands (Dirac cones) with little or no doping.\cite{Hicks_JPhysD_12}  The band structure of a graphene film with 0.2ML interface N-content is dramatically different as seen in Fig.~\ref{F:ARPES}(b) and (c).  When interfacial nitrogen is present, a band-gap has developed.  Figure \ref{F:ARPES}(d) shows energy distribution curves (EDC) through the $K$-point at $k_y\!=\!0$ from (a), (b), and (c). The EDC for the clean surface is peaked near the $E_F$ (the small intensity at 0.5eV is from a faint cone from the layer below rotated relative to the top layer, typical of C-face graphene).\cite{Sprinkle_PRL_09} Unlike pristine graphene, the peak in the EDCs from the nitrogen-seeded samples are shifted to higher BEs indicating a valence band maximum corresponding to an energy gap.  Figure \ref{F:ARPES}(d) shows that the 3-layer N-seeded sample has a 0.7 eV gap (depending on the position of the conduction band minimum), while the 8-layer film has a smaller 0.45 eV gap (a 4-layer film, not shown, confirms this trend).  The C1s spectra from clean and N-seeded samples show that the valence band spectra cannot be due to an energy shift caused by band bending changes in different samples [see supplement]. We also note that, the effective  velocity, $\tilde{v}$, derived from a linear approximation to $E(k)$, is reduced compared to $v_F$, consistent with the opening of a band-gap.  Within a small $k_y$-range around the Dirac point ($\Delta k_y\!=\!\pm0.15\text{\AA}^{-1}$), both the 3- and 8-layer samples have the same $\tilde{v}$ within the error imposed by the ARPES broadening ($\tilde{v}\!=\!0.8\!\pm0.05\!\times\!10^6$m/sec).

Compared to pristine graphene, the nitrogen-seeded graphene $\pi$ bands are broader in $k$. The $\Delta k$ momentum broadening is $\sim\!0.25\text{\AA}^{-1}$ FWHM  (independent of layer thickness) [see supplemental material].  A large part of the EDC energy broadening and the momentum broadening is due to the corrugation of the graphene surface [the corrugation is demonstrated in STM results presented below].\cite{Park_NLet_09,Knox_PRB_11} Small modulations in the local graphene height cause a local angular variation in the surface normal.  Since the surface normal determines the orientation of the graphene Brillouin zone, the corrugated surface leads to local $k_x$ and $k_y$ shifts in the $K$-point.  This leads to an ARPES image that is an area average of a distribution of parabolic cuts through Dirac cones from local tilted graphene resulting in an $E$- and $k$-broadened spectra.  While this explains part of the broad intensity distribution between the valence band maximum and $E_F$.  A more detailed STM analysis suggest how the gap forms and an additional broadening of the EDCs.


\begin{figure}[tb]
\includegraphics[width=7.5cm]{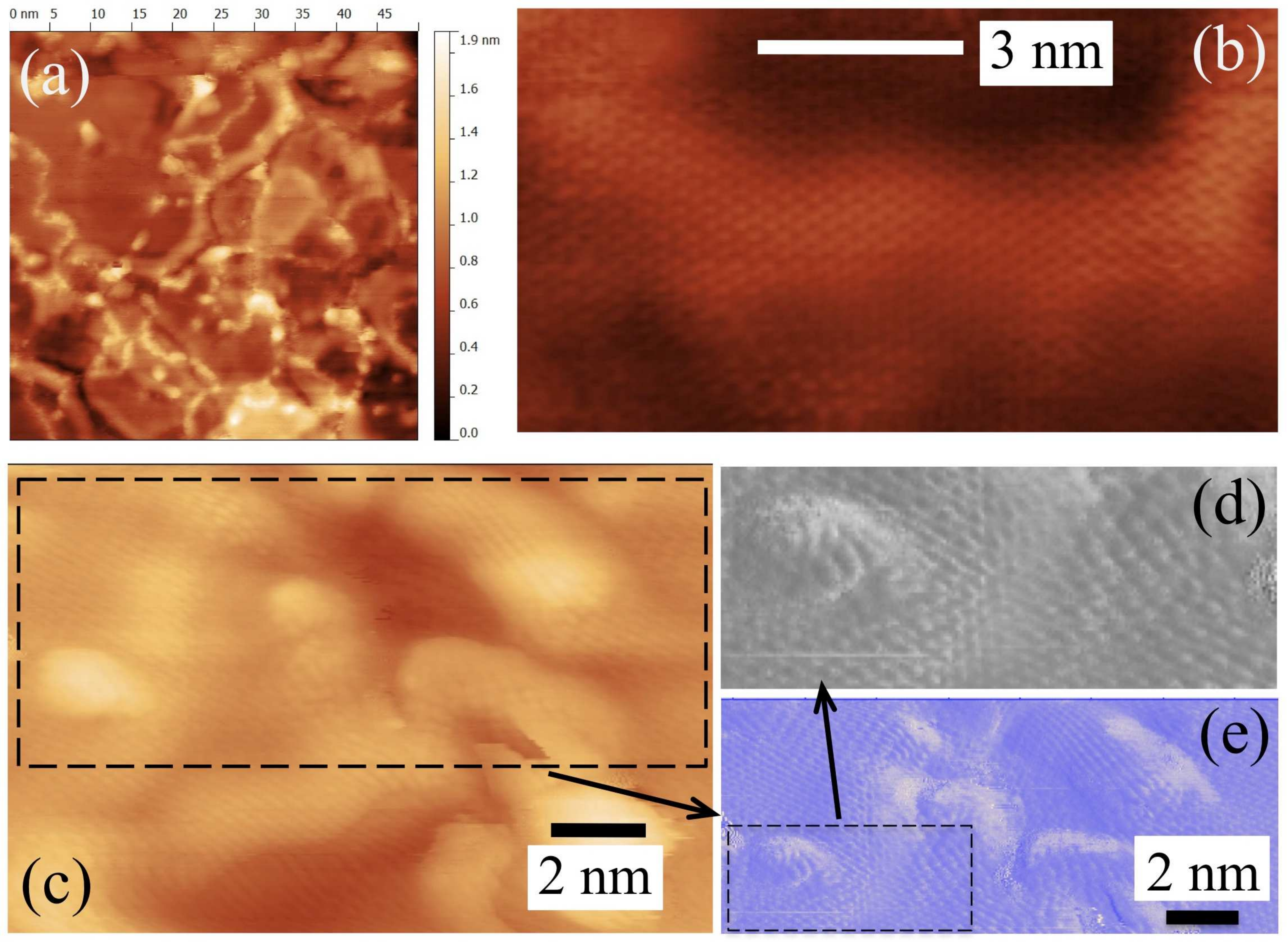}
\caption{(a) $50\times50\text{nm}^2$ STM image of a 3-layer graphene film grown from a 0.2ML nitrogen-seeded SiC$(000\bar{1})$ surface (bias voltage, $V_B\!=\!0.1$V).  The image shows a pattern of folds. (b) A topographic image of a fold showing the graphene lattice ($V_B\!=\!0.1$V). (c) A $10\!\times\!10$nm view of the graphene folds. (d) A magnified $dI/dV$ images of folded graphene showing that the graphene is continuous over the folds and in the valleys.  (e) A $dI/dV$ image of a region in (c).  The image shows that the tops of the folds are bright in the $dI/dV$ image indicating a high density of states.} \label{F:STM}
\end{figure}

STM images of the nitrogenated graphene show that the films are not flat like typical C-face graphene films.\cite{Miller_Nphy_10}  Figure \ref{F:STM}(a) show the highly buckled topography. A close up view of a buckled regions in Fig.~\ref{F:STM}(b) shows that the fold is made of intact graphene.  Note that there are no indications of nitrogen inclusions like those seen in CVD and plasma created nitrogen.\cite{Zhao_Sci_11,Joucken_PRB_12}  The folds are $\sim\!1$nm (no more than 2nm) high and 2-4nm wide giving a radius of curvature of 1-2nm. These folds meander and can extend up to 25nm but are typically 5nm long.  Figure \ref{F:STM}(c) shows an STM topographic image of a region with several folds.  A $dI/dV$ image in Fig.~\ref{F:STM}(d) of a small region between folds confirms that the graphene is continuous.  We note that this folding is not a property of epitaxial graphene.  The RMS roughness is less than 0.005nm for graphene films without nitrogen.\cite{Hass_JPhyCM_08}

It is important to point out that the 2-4nm width of these folds is consistent with the $0.25\text{\AA}^{-1}$ $\Delta k$ broadening of the ARPES spectrum.  In ARPES, $\Delta k\!\sim\!2\pi/L$ where $L$ is the average coherent domain size.  This allows us to estimate $L\!\sim\!2.5\text{nm}$ from the ARPES broadening. In addition to the scale of the finite size domains measured by ARPES, the mean separation between $N_P$ nitrogen atoms, determined by XPS, is 1.5nm; only slightly smaller than the width of the folds determined by STM.

That fact that three different techniques measure similar length scales is not a coincidence. A model consistent with these results is shown in Fig.~\ref{F:N2_Growth}.  During the high temperature growth, Si evaporates leaving a carbon rich surface with N-impurities. As the graphene layer crystalizes from this film at high temperature, N atoms in high energy interstitial graphene sites are expelled from the growing graphene.  Some of these nitrogen desorb while the others re-bind to the SiC as either $N_S$ or $N_P$ nitrogen [see Fig.~\ref{F:N2_Growth}(a)]. The interfacial nitrogen atoms ($N_P$) act as pinning sites that lock the graphene to the SiC. This is supported by a large D-peak in Raman spectra consistent with a significant concentration of $\text{sp}^3$-like bonds that would be associated with graphene-N bonds [see supplemental material]. These $\text{sp}^3$ bonded C-N-SiC sites constrain the graphene to a length scale related to the starting nitrogen concentration.  Since the N-constrained graphene is no longer commensurate with the SiC, a strain develops in the film that forces the graphene to buckle.
The length scale between folds is set by the strain field induced by the interfacial nitrogen concentration. As the next graphene layer forms, Si evaporation causes the SiC interface to recede along with the interfacial nitrogen as shown in Figs.~\ref{F:N2_Growth}(b) and (c). As each new graphene layer forms the process is repeated leaving a stack of $\pi$-bonded graphene layers with only the last layer nitrogen bonded to the SiC. The buckling period remains determined by the nitrogen concentration and not the number of layers (confirmed by the layer independent ARPES broadening). As Fig.~\ref{F:N2_Growth} shows, the buckling amplitude is expected to reduce in layers farther from the SiC as previously shown by STM.\cite{Riedl_PRB_07}


It is these folds that are responsible for the energy gap and not nitrogen impurities in the graphene. Because of the short electron mean free path, the ARPES only measures the graphene band structure in the top 2-3 layers where the XPS shows no measurable nitrogen concentration. This means that the measured band-gap must be related to the graphene folds.  The size of the ARPES gap and its dependence on the layer thickness point to a finite size effect gap caused by either strain confined domain boundaries or by a quasi periodic strain field.

The boundary between folds is a highly strained region of graphene that could confine the graphene wave functions between folds. The confinement band-gap expected for a graphene ribbon with width $w\!=\!1\!-\!2$ nm is $E_g\sim\!1\text{eV-nm}/w\!=\!1.0\!-\!0.5$ eV,\cite{Wakabayashi_STAM_10,Son_PRL_06} close to the value measured.  The fact that a distribution of confined folds are part of a continuous sheet of graphene would also explain the ARPES intensity seen within the gap.  A model of bent graphene confined by flat graphene sheets predicts a significant density of states (DOS) in the gap due to boundary resonant states.\cite{Hicks_NatureP_13}  The finite size model also explains the smaller gap in thicker films even though the domain size is the same.  As the number of layers increases, the amplitude of the buckles reduces, thus reducing the strain at the bends. This makes the confinement boundary less well defined; effectively increasing the confined region and reducing the gap.  Because of the fold width distribution, the area-averaged ARPES would have a distribution of band-gaps resulting in an increased intensity between the top of the valance band and $E_F$ as observed.

Periodic strain fields are also expected to open a band-gap.\cite{Low_PRB_11}  Figure \ref{F:STM}(e) shows a  $dI/dV$ image of several fold ridges that show the typical increased DOS predicted for strained graphene\cite{Pereira_PRL_10} and observed in STM on graphene nano-bubbles.\cite{Levy_Science_10}  While the nitrogen induced folds are not strictly periodic, the folds have an average separation of about 2-4 nm.  This is a very small period that has the potential to open a large gap.  Whether strain or electron confinement (or both) are responsible for the observed gap remains to be determined.


We have shown that a partial nitrogen monolayer, pre-grown on the SiC$(000\bar{1})$ surface, can be used to form an undulating graphene layer with a band-gap $\sim\!0.7$eV.  The nitrogen binds the growing graphene to the SiC interface forming 1-2nm high ridges in the graphene. There is no nitrogen intercalation between graphene sheets nor is there evidence of nitrogen inclusions that can reduce graphene's mobility.\cite{Wei_NanoLet_09,Jin_ACSNano_11}
Because the initial nitrogen-carbon bond is stable to very high temperatures, the pre-grown layer can be patterned to produce locally strained graphene.  This semiconducting form of graphene would then be seamlessly connected to metallic graphene grown from the non-nitrogenated SiC.  It offers a potential way to produce all graphene semiconductor-metal junctions.

\begin{acknowledgments}
This research was supported by the NSF under Grants No. DMR-1206793, -1206655, and -1206256. Additional support is also acknowledged from the NSF DMR-1005880 and the W.M. Keck Foundation.  We thank A. Savu for growing the graphene and Z. Dromsky for taking Raman spectra.  We wish to acknowledge the SOLEIL synchrotron and the Cassiop\'{e}e beam line staff. We wish to thank N. Guisinger for his assistance in the STM measurements performed at the Center for Nanoscale Materials supported by the U. S. DOE, Office of Science, Office of Basic Energy Sciences, under Contract No. DE-AC02-06CH11357.  We thank Profs. J. Williams and S. Dhar for useful discussions concerning the NO process.
\end{acknowledgments}


\noindent \textbf{Experimental Method}

\noindent The substrates used in these studies were n-doped (nitrogen) 4H-SiC.  All the graphitized samples were grown in a closed RF induction furnace using the Confined Silicon Sublimation method.\cite{WaltPNAS}  The samples were transported in air before introduction into either the XPS, STM, or the ARPES UHV chamber.  ARPES and photon energy dependent XPS measurements were made at the Cassiop\'{e}e beamline at the SOLEIL synchrotron in Gif sur Yvette. The high resolution Cassiop\'{e}e beamline has a total measured instrument resolution $\Delta E\!<\!12$meV using a Scienta R4000 detector with a $\pm 15^{\circ}$ acceptance  at $\hbar \omega\!=\!36$ eV.

To produce the initial nitrogen surface layers the SiC substrates were RCA cleaned.  The samples were loaded into a 900\degC furnace under a 500 sccm Ar flow and heated to 1175\degC over a 1hr ramp.  The sample is then kept at 1175\degC for 2 hrs (for a  0.3ML nitrogen coverage) with a 500sccm NO flow.\cite{Liu_IEEE_13}  The sample is then cooled to 900\degC under a 500 sccm Ar flow and unloaded from furnace. Oxide grown through this anneal is removed by HF immediately before graphene growth. These C-face nitrogen-seeded surfaces were then heated in a closed graphite crucible in an RF vacuum furnace to 1450\degC to produce the graphene films.  This growth temperature is slightly higher that the desorption temperature of nitrogen as discussed in the supplement.

The nitrogen coverage, $N_{N}$, is estimated from the ratio of the N1s to Si 2p intensities $N_{N}=\dfrac {I_{N}} {I_{Si}}\dfrac{\sigma_{Si}}{\sigma_{N}}n_{Si}\lambda$, where $I_{N}$ and ${I_{Si}}$ are the N1s and Si 2p XPS
intensities, $\sigma_{N}$ and $\sigma_{Si}$ are the photoionization cross sections  of N and Si.\cite{Yeh_ADND_85,Yeh_Book_93} $n_{Si}\!=\!4.8\!\times\!10^{22}/\text{cm}^3$ is atomic density of Si in SiC. For this work, we use a mean free path in SiC of $\lambda=2.2$nm at $1486$eV.

\noindent \textbf{Supporting Information}

Pre-growth Nitrogen preparation, Nitrogen Site assignments, ARPES band structure analysis, and Raman analysis are available in the supporting information. This material is available free of charge via the Internet at http://pubs.acs.org.

\end{document}